\documentclass[aps,prl,twocolumn,showpacs,superscriptaddress,floatfix]{revtex4}
\usepackage{pstricks,pst-node,pst-text,pst-3d,pst-plot}
\usepackage{graphicx}

\usepackage{graphicx}
\usepackage[english]{babel}
\usepackage{bm}    
\usepackage{eepic} 

\def\smfrac#1#2{{\textstyle\frac{#1}{#2}}}

\def\reff#1{(\ref{#1})}
\newcommand{\be}{\begin{equation}}
\newcommand{\ee}{\end{equation}}

\def\spose#1{\hbox to 0pt{#1\hss}}
\def\ltapprox{\mathrel{\spose{\lower 3pt\hbox{$\mathchar"218$}}
 \raise 2.0pt\hbox{$\mathchar"13C$}}}
\def\gtapprox{\mathrel{\spose{\lower 3pt\hbox{$\mathchar"218$}}
 \raise 2.0pt\hbox{$\mathchar"13E$}}}

\newcommand{\scre}{{\cal E}}

\newcommand{\scrm}{{\cal M}}

\newcommand{\Ghat}{{\widehat{G}}}
\newcommand{\Ehat}{{\widehat{E}}}
\newcommand{\Gtilde}{{\widetilde{G}}}

\newsavebox{\fancyplusb}

\savebox{\fancyplusb}(2,2){
\thinlines
\put(0,-1){\line(0,1){2}}
\put(-1,-0.5){\line(0,1){1}}
\put(-1,0){\line(1,0){2}}
\put(1,-0.5){\line(0,1){1}}
\put(-0.5,-1){\line(1,0){1}}
\put(-0.5,1){\line(1,0){1}}
}

\begin{document}

\title{Finite-temperature phase transition in a class of
       4-state Potts antiferromagnets}


\author{Youjin Deng}
\email{yjdeng@ustc.edu.cn}
\author{Yuan Huang}
\email{huangy22@mail.ustc.edu.cn}
\affiliation{Hefei National Laboratory for Physical Sciences at Microscale
   and Department of Modern Physics,
   University of Science and Technology of China,
   Hefei, Anhui 230026, China}
\author{Jesper Lykke Jacobsen}
\email{jacobsen@lpt.ens.fr}
\affiliation{Laboratoire de Physique Th\'eorique,
   \'Ecole Normale Sup\'erieure, 24 rue Lhomond, 75231 Paris, France}
\affiliation{Universit\'e Pierre et Marie Curie,
   4 place Jussieu, 75252 Paris, France}
\author{Jes\'us Salas}
\email{jsalas@math.uc3m.es}
\affiliation{
  Gregorio Mill\'an Institute,
  Universidad Carlos III de Madrid,
  28911 Legan\'es, Spain}
\author{Alan D. Sokal}
\email{sokal@nyu.edu}
\affiliation{Department of Physics, New York University,
      4 Washington Place, New York, NY 10003, USA}
\affiliation{Department of Mathematics, University College London,
      London WC1E 6BT, UK}

\date{August 8, 2011}

\begin{abstract}
We argue that
the 4-state Potts antiferromagnet has
a finite-temperature phase transition
on any Eulerian plane triangulation in which one sublattice
consists
of vertices of degree 4.
We furthermore predict the universality class of this transition.
We then present transfer-matrix and Monte Carlo data
confirming these predictions for the cases of the
union-jack and bisected hexagonal lattices.
\end{abstract}



\maketitle

The $q$-state Potts model \cite{Potts_52,Wu_82+84}
plays an important role in the theory of critical phenomena,
especially in two dimensions \cite{Baxter_book,Nienhuis_84,DiFrancesco_97},
and has applications to various condensed-matter systems \cite{Wu_82+84}.
Ferromagnetic Potts models are by now fairly well understood,
thanks to universality;
but the behavior of antiferromagnetic Potts models
depends strongly on the microscopic lattice structure,
so that many basic questions must be investigated case-by-case:
Is there a phase transition at finite temperature, and if so, of what order?
What is the nature of the low-temperature phase(s)?
If there is a critical point, what are the critical exponents and the
universality classes?
Can these exponents be understood (for two-dimensional models)
in terms of conformal field theory \cite{DiFrancesco_97}?

One expects that for each lattice ${\cal L}$ there
exists a value $q_c({\cal L})$ [possibly noninteger]
such that for $q > q_c({\cal L})$  the model has exponential decay
of correlations at all temperatures including zero,
while for $q = q_c({\cal L})$ the model has a zero-temperature critical point.
The first task, for any lattice, is thus to determine $q_c$.

Some two-dimensional (2D) antiferromagnetic models at zero temperature
can be mapped
exactly
onto a ``height''
model (in general vector-valued) \cite{Salas_98,Jacobsen_09}.
Since the height model must either be in a ``smooth'' (ordered)
or ``rough'' (massless) phase,
the corresponding zero-temperature spin model must either be
ordered or critical, never disordered.
Experience tells us that the most common case is criticality
\cite{height_rep_exceptions}.
The long-distance behavior is then that of a massless Gaussian
with some
({\em a~priori}\/  
unknown) ``stiffness matrix'' ${\bf K} > 0$.
The critical operators can be identified via the height mapping,
and the corresponding critical exponents can be predicted in terms
of ${\bf K}$.
Height representations thus provide a means for recovering
a sort of universality for some (but not all) antiferromagnetic models
and for understanding their critical behavior
in terms of conformal field theory. 

In particular, when the $q$-state zero-temperature Potts antiferromagnet
on a
2D lattice ${\cal L}$ admits a height representation,
one ordinarily expects that $q = q_c({\cal L})$.
This prediction is confirmed in most heretofore-studied cases:
3-state square-lattice \cite{Nijs_82,Kolafa_84,Burton_Henley_97,Salas_98},
3-state kagome \cite{Huse_92,Kondev_96},
4-state triangular \cite{Moore_00},
and 4-state on the line graph
of the square lattice
\cite{Kondev_95,Kondev_96}.
The only known exceptions
are the triangular Ising antiferromagnet \cite{note_TRI_q=2}
and the 3-state model on the diced lattice \cite{Kotecky-Salas-Sokal}.

Moore and Newman \cite{Moore_00}
observed
that the height mapping employed for the
4-state Potts antiferromagnet on the triangular lattice
carries over unchanged to any Eulerian plane triangulation
(a graph is called Eulerian if all vertices have even degree;
it is called a triangulation if all faces are triangles).
One therefore expects naively that $q_c = 4$
for every (periodic) Eulerian plane triangulation.

Here we will present analytic arguments suggesting that
this naive prediction is {\em false}\/ for an infinite class of
Eulerian plane triangulations, namely those in which one sublattice
consists entirely of vertices of degree 4.
More precisely, we predict that on these lattices
the 4-state Potts antiferromagnet has a phase transition
at \hbox{{\em finite}\/} temperature (so that $q_c > 4$);
we shall also predict the universality class of this transition.
We will conclude by presenting transfer-matrix and Monte Carlo data
confirming these predictions for the cases of the
union-jack [$D(4,8^2)$] and bisected hexagonal [$D(4,6,12)$] lattices.

\begin{figure}
\begin{center}
\includegraphics[width=0.47\columnwidth]{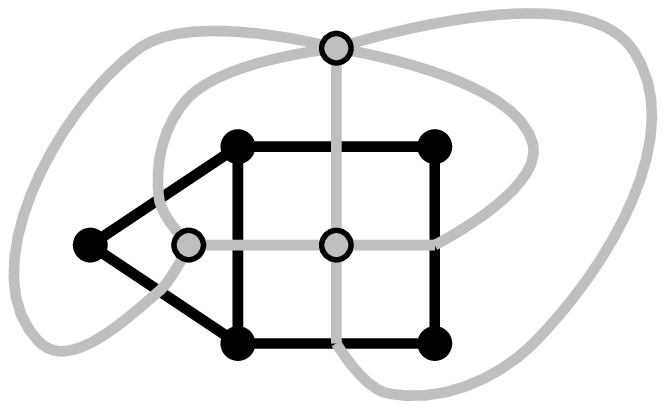}
\hspace*{2mm}
\includegraphics[width=0.47\columnwidth]{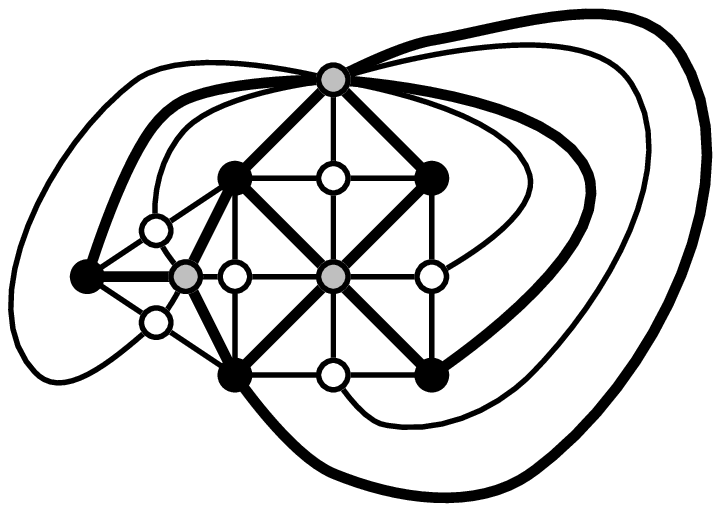} \\
\end{center}
\vspace*{-5mm}
\caption{
   (a) A dual pair $G$ (black dots and lines)
          and $G^*$ (gray dots and lines).
   (b) The quadrangulation $\Ghat$ (\hbox{black/gray} dots and thick lines)
          and
          triangulation $\Gtilde$ (all dots and lines).
}
\label{fig1}
\end{figure}

\paragraph{Exact identities.}
Let $G=(V,E)$ and $G^* = (V^*, E^*)$ be a dual pair
of connected
graphs embedded in the plane (Fig.~\ref{fig1}a).
Then define $\Ghat = (V \cup V^*, \Ehat)$
to be the graph with vertex set $V \cup V^*$
and edges $ij$
whenever $i \in V$ lies on the boundary of the face of $G$
that contains $j \in V^*$ (Fig.~\ref{fig1}b).
The graph $\Ghat$ is a plane quadrangulation:
on each face of $\Ghat$, one pair of diametrically opposite vertices
corresponds to an edge $e \in E$
and the other pair corresponds to the dual edge $e^* \in E^*$.
In fact, $\Ghat$ is nothing other than the dual
of the medial graph $\scrm(G) = \scrm(G^*)$ \cite{Godsil-Royle}.
Conversely, every plane quadrangulation $\Ghat$
arises via this construction from some pair $G,G^*$.

%
%

Now let $\Gtilde$ be the graph obtained from $\Ghat$
by adjoining a new vertex in each face of $\Ghat$
and four new edges connecting this new vertex to the four corners of the face
(Fig.~\ref{fig1}b).
This graph $\Gtilde$ is an Eulerian plane triangulation,
with vertex tripartition $V \cup V^* \cup C$
where $C$ is the set consisting of the ``new'' degree-4 vertices.
Conversely, every Eulerian plane triangulation
in which one sublattice consists of degree-4 vertices
arises
in this way.

We will show elsewhere \cite{potts-rsos}
that the 4-state Potts antiferromagnet at zero temperature
(= 4-coloring model) on $\Gtilde$
can be mapped exactly onto the 9-state Potts {\em ferro}\/magnet
on $G$ and $G^*$:
\begin{equation}
   Z_{\Gtilde}(4,-1)  \:=\:  4 \cdot 3^{-|V|} \, Z_G(9,3)
                     \:=\:  4 \cdot 3^{-|V^*|} \, Z_{G^*}(9,3)
   \;
   \label{eq.prop_4color}
\end{equation}
where $Z_G(q,v)$ denotes the Potts-model partition function
with $v = e^J - 1$ ($J = $ nearest-neighbor coupling).
The proof passes either via an RSOS model on $\Ghat$
or via a completely packed loop model on $\scrm(G)$.

\paragraph{Height representation \cite{Moore_00}.}
Consider the 4-coloring model
on an Eulerian plane triangulation $\Theta$.
We can orient the edges of $\Theta$ such that
the three edges around each face define a cycle
(clockwise on one sublattice of $\Theta^*$ and
counterclockwise on the other).
Let ${\bf e}_0, {\bf e}_1, {\bf e}_2$ be unit vectors
at angles $0, 2\pi/3, 4\pi/3$ in the plane.
Then, to any proper 4-coloring $\sigma$ of the vertices of $\Theta$,
we assign heights ${\bf h}_i$ in the triangular lattice
such that ${\bf h}_j - {\bf h}_i = {\bf e}_0, {\bf e}_1, {\bf e}_2$
on an oriented edge $\vec{ij}$ according as
$\{\sigma_i,\sigma_j\} = \{1,2\}$ or $\{3,4\}$,
$\{1,3\}$ or $\{2,4\}$, $\{1,4\}$ or $\{2,3\}$.

\paragraph{Phase transition and universality class.}
Let now $G$, $G^*$ and $\Gtilde$ be infinite regular lattices.
Conformal field theory \cite{DiFrancesco_97}
tells us that a $q$-state Potts ferromagnet with $q > 4$
cannot have a critical point.
Therefore the 9-state Potts ferromagnet in \reff{eq.prop_4color}
is noncritical, suggesting that the
4-state Potts antiferromagnet at zero temperature on $\Gtilde$
is also noncritical \cite{note_criticality}.
But since this model has a height representation,
it cannot be disordered;  therefore it must be ordered.
It follows that the 4-state Potts antiferromagnet on $\Gtilde$
has an order-disorder transition (whether first-order or second-order)
at finite temperature.

We can also understand the type of order in the
4-coloring model on $\Gtilde$,
and hence the universality class of the order-disorder transition
in case it is second-order.
If the lattice $G$ is self-dual,
the point $(q,v) = (9,3)$ lies on the self-dual curve $v = \sqrt{q}$,
which is expected to be the locus of first-order transitions;
so there are phases in which $G$ is ordered and $G^*$ is disordered,
and vice versa (nine of each).
We therefore predict that the 4-coloring model on $\Gtilde$
has phases in which the sublattice $V$ is ordered in
one of the four possible directions while $V^*$ and $C$ are disordered,
and the same with $V$ and $V^*$ interchanged.
The symmetry is $S_4 \times Z_2$,
so we expect that the transition is in the universality class
of a 4-state Potts model plus an Ising model (decoupled).
On the other hand, if $G$ is not self-dual,
then we expect (barring a fluke) that $(q,v) = (9,3)$
does {\em not}\/ lie on a phase-transition curve;
hence one of the lattices $G$, $G^*$ will be ordered (say, $G$)
while the other is disordered.
In this case we predict that the 4-coloring model on $\Gtilde$
has phases in which the sublattice $V$ is ordered in
one of the four possible directions while $V^*$ and $C$ are disordered.
The symmetry is $S_4$,
and the universality class is that of a 4-state Potts model.

We recall that the central charge $c$ and
magnetic and thermal exponents $X_m,X_t$
are $(c,X_m,X_t) = (\smfrac{1}{2},\smfrac{1}{8},1)$ for the Ising model
and $(1,\smfrac{1}{8},\smfrac{1}{2})$ for the 4-state Potts model.

%
%

\begin{figure}
\begin{center}
\includegraphics[width=0.439\columnwidth]{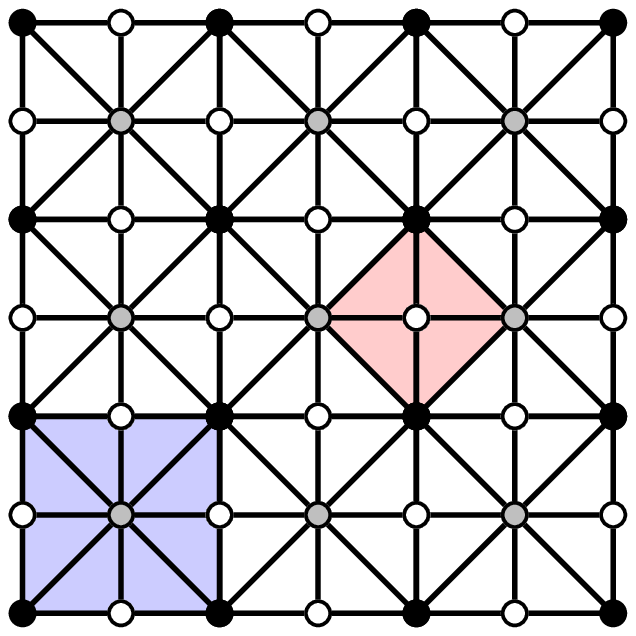}
\hspace*{2mm}
\includegraphics[width=0.501\columnwidth]{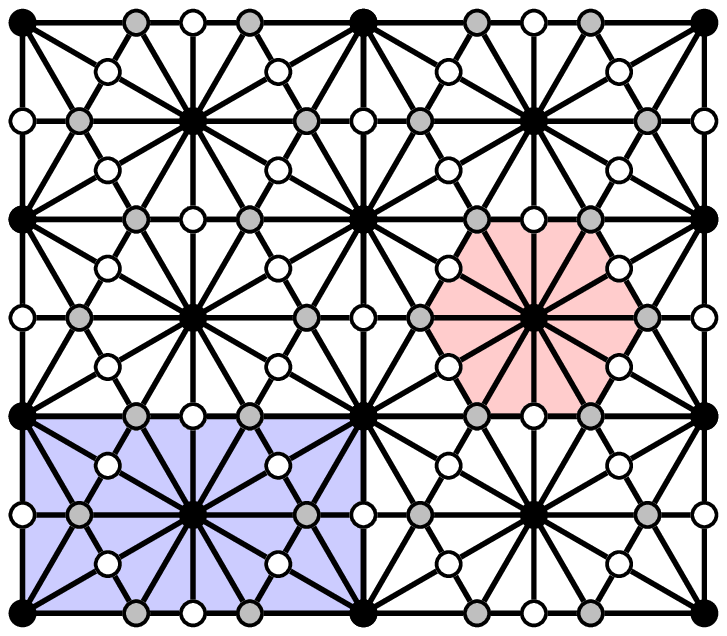}
\end{center}
\vspace*{-4mm}
\caption{
   (a) Union-jack lattice, $L=6$.
   (b) Bisected hexagonal lattice, $L=8$.
   The shaded areas show the minimal unit cells (pink)
   and the rectangular unit cells used in the row-to-row transfer-matrix
   computations (blue).
   The tripartition of the vertex set is shown in black/gray/white
   as in Fig.~\ref{fig1}.
}
\label{fig:lattices}
\end{figure}

\begin{figure}[t]
\vspace*{-1mm}
\begin{center}
\includegraphics[width=\columnwidth]{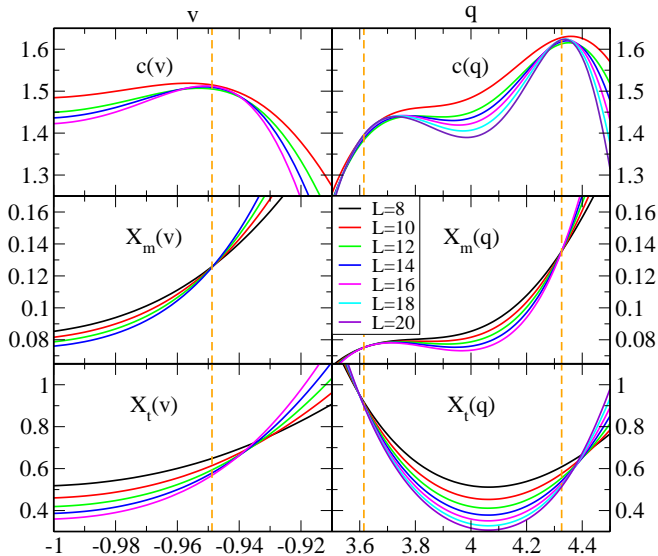}
\end{center}
\vspace*{-4mm}
\caption{
  Estimates for central charge $c$ and critical exponents $X_m,X_t$
  for the UJ lattice, as a function of $v$ at $q=4$ (left)
  and as a function of $q$ at $v=-1$ (right).
  Dashed vertical lines indicate our best FSS estimates of
  $v_c$, $q_0$ and $q_c$.
  Fits used Eqs.~\reff{eq.FSS.c}/\reff{eq.FSS.X}
  with $o(1/L^2)$ replaced by $A/L^4$,
  for three (resp.\ two) consecutive values of $L$.
}
\label{fig_TM_data_UJ}
\end{figure}

\paragraph{Union-jack (UJ) lattice.}
$\!\!\!$The simplest self-dual case is  $G = G^* = \Ghat = $ square lattice
and $\Gtilde = $ \hbox{union-jack lattice} (Fig.~\ref{fig:lattices}a).
We computed transfer matrices with
fully
periodic boundary conditions
for even widths $L \le 20$ ($v=-1$) \cite{note_Jacobsen_10}
and $L \le 16$ (general $v$).
Estimates of
$c,X_m,X_t$
were extracted from the free energy $f_L$ and free-energy gaps $\Delta f_L$
via \cite{DiFrancesco_97}
\begin{eqnarray}
   f_L         & = &  f_\infty \,-\, \pi c/(6L^2) \,+\, o(1/L^2)
      \label{eq.FSS.c}   \\
   \Delta f_L  & = &  2\pi X/L^2 \,+\, o(1/L^2)
      \label{eq.FSS.X}
\end{eqnarray}
Fig.~\ref{fig_TM_data_UJ} (upper left)
shows estimates of $c(v)$ at $q=4$.
\hbox{The maximum} at $v \!\approx\! -0.95$
\hbox{indicates the
transition:}
\hbox{finite-size} scaling (FSS)
yields
$v_c = -0.944(5)$ and $c = 1.510(5)$,
in agreement with our prediction $c = 1 + \smfrac{1}{2} = \smfrac{3}{2}$.
The crossings of $X_m(v)$ and $X_t(v)$ yield
$v_c = -0.9488(3)$, $X_m = 0.1255(6)$ and $X_t = 0.51(2)$,
in agreement with $X_m = \smfrac{1}{8}$ and $X_t = \smfrac{1}{2}$
\cite{note_Xt}.

A similar plot for $c(q)$ at $v=-1$
shows the
lattice-independent
Berker--Kadanoff phase [$c = 1 - 6(t-1)^2/t$ with $q = 4 \cos^2 (\pi/t)$]
for $0 \le q < q_0$
and a noncritical phase for $q_0 < q < q_c$.
The maxima of $c(q)$
[Fig.~\ref{fig_TM_data_UJ}, upper right]
yield the estimates $q_0 = 3.63(2)$, $c = 1.43(1)$
and $q_c = 4.330(5)$, $c = 1.63(1)$.
The crossings of $X_m(q)$ and $X_t(q)$ yield
$q_0 = 3.616(6)$, $X_m = 0.0751(3)$, $X_t = 0.88(2)$
and $q_c = 4.326(5)$, $X_m = 0.134(2)$, $X_t = 0.52(3)$.

The data for $q_0$ are consistent with
$q_0 = B_{10} = (5 + \sqrt{5})/2 \approx 3.61803$ \cite{note_Beraha}
and $c= 7/5$, $X_m = 3/40$, $X_t = 7/8$;
the underlying conformal field theory could be
a pair of $m = 4$ minimal models \cite{note_m=4}.

Concerning $q_c$, we have seen that at $(q,v) = (4,v_c)$
the critical behavior is that of a
4-state Potts model plus an Ising model (decoupled),
and it is compelling to think that this behavior will persist along
a curve in the $(q,v)$-plane passing through $(q,v) = (q_c,-1)$.
However, it is possible that $(q_c,-1)$ might be
the endpoint of this curve, in which case the model could be
driven there to some
sort of
multicritical behavior:
for instance, a 4-state Potts model plus a {\em tricritical}\/
Ising model (decoupled), which would have
$c = 1 + \smfrac{7}{10} = \smfrac{17}{10}$
and $X_m = X_{1/2,0} = 21/160 = 0.13125$ \cite{note_m=4}.
Alternatively, the critical exponents along the transition curve
may vary continuously with $q$.

%

%
%

\begin{figure}[t]
\vspace*{-1mm}
\begin{center}
\includegraphics[width=0.9\columnwidth]{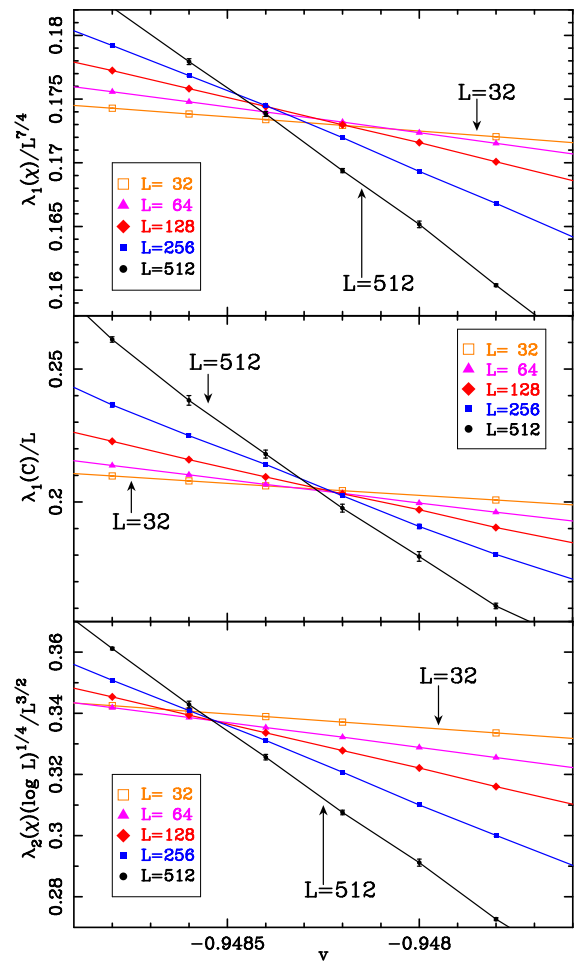}
\end{center}
\vspace*{-6mm}
\caption{
  Monte Carlo
  data for the UJ lattice \hbox{at $q=4$.}
  (a) Leading susceptibility eigenvalue $\lambda_1(\chi)$ divided by $L^{7/4}$.
  (b) Leading specific-heat eigenvalue $\lambda_1(C)$ divided by $L$.
  (c) Second susceptibility eigenvalue $\lambda_2(\chi)$ divided by
       $L^{3/2} (\log L)^{-1/4}$.
  Lines are meant only to guide the eye.
}
\label{fig_UJ_MC_data}
\end{figure}

%
%

We also simulated the $q=4$ model,
using a \hbox{cluster} Monte Carlo (MC) algorithm,
on periodic $L \times L$ lattices with $8 \le L \le 512$.
We measured the sublattice magnetizations $\scrm_A,\scrm_B,\scrm_C$,
the nearest-neighbor correlations $\scre_{AB}, \scre_{AC}, \scre_{BC}$
and the next-nearest-neighbor correlations
$\scre_{AA}, \scre_{BB}, \scre_{CC}$.
We then computed the $3 \times 3$ sublattice susceptibility matrix
and the $6 \times 6$ specific-heat matrix;
from their eigenvalues we can extract
the magnetic and thermal critical exponents.
%
%
The leading susceptibility eigenvalue diverges with the predicted exponent
$\gamma/\nu = 2 - 2X_m = 7/4$ (Fig.~\ref{fig_UJ_MC_data}a),
and FSS analysis yields the estimate $v_c = -0.9485(1)$.
Likewise,
the leading specific-heat eigenvalue diverges with exponent
$\alpha/\nu = 2 - 2X_t = 1$ \cite{note_Xt} (Fig.~\ref{fig_UJ_MC_data}b),
and FSS analysis yields the estimate $v_c = -0.9483(2)$.
It is curious that we do not see here the multiplicative logarithms
that are expected \cite{Salas-Sokal_4-state} for the 4-state Potts model.
The second susceptibility eigenvalue diverges as $L^{3/2}$,
probably with a multiplicative logarithm (Fig.~\ref{fig_UJ_MC_data}c),
while the second specific-heat eigenvalue tends to a finite constant;
we have no theoretical understanding of these behaviors.

A transition in this model
was recently predicted by \hbox{Chen {\em et al.}\@$\!$}
\cite{Chen_11},
who found $v_c = -0.9477(5)$ by a \hbox{tensor} renormalization-group method;
they also gave an entropy-counting argument predicting the type of order.
However, in their approximation the specific heat was non-divergent,
exhibiting a jump discontinuity.

\paragraph{Bisected hexagonal (BH) lattice.}
The simplest non-self-dual case is $G = $ triangular lattice and
$G^* = $ hexagonal lattice,
yielding $\Ghat = $ diced lattice
and $\Gtilde = $ bisected hexagonal lattice (Fig.~\ref{fig:lattices}b).
We computed transfer matrices with fully periodic boundary conditions
for the same widths as for the UJ lattice,
except that $L$ must now be a multiple of 4
to be compatible with the periodic boundary conditions
(see Fig.~\ref{fig:lattices}b).
FSS analysis of $c(v)$ at $q=4$
yields the estimates $v_c = -0.8281(1)$ and $c = 1.000(5)$,
in agreement with our prediction $c = 1$.
The crossing of $X_m(v)$ and the minimum of $X_t(v)$ yield
$v_c = -0.8280(1)$, $X_m = 0.15(1)$ and $X_t = 0.65(10)$,
which are compatible with $X_m = \smfrac{1}{8}$ and $X_t = \smfrac{1}{2}$
although rather imprecise.

The maximum of $c(q)$ yields
$q_c = 5.395 (10)$, $c = 1.20(5)$.
The crossing of $X_m(q)$ and the minimum of $X_t(q)$ yield
$q_c = 5.397(5)$, $X_m = 0.15(1)$, $X_t = 0.6(1)$.

We also did MC simulations for $q=4$ and $8 \le L \le 512$.
The leading susceptibility eigenvalue diverges as expected as $L^{7/4}$
[possibly with a multiplicative $(\log L)^{-1/8}$]
and yields $v_c = -0.828066(4)$.
The leading specific-heat eigenvalue
is compatible with the 4-state Potts behavior $L (\log L)^{-3/2}$.

A transition in this model was also conjectured in \cite{Chen_11}.

Our result $q_c > 5$ suggests that there will be
a finite-temperature transition also in the {\em 5-state}\/ model.
Quite surprisingly, we find this transition to be {\em second-order}\/,
despite the absence of an obvious universality class (since $q>4$);
however, it is also conceivable that the transition
is weakly first-order,
with a correlation length that is finite but very large.
Preliminary
results from transfer matrices are
$v'_c = -0.9513(1)$, $c = 1.17(5)$, $X_m = 0.16(1)$, $X_t = 0.56(4)$.
Preliminary MC results are
$v'_c = -0.95132(2)$, $X_m = 0.113(4)$, $X_t = 0.495(5)$.
More detailed data will be reported separately \cite{all_of_us_in_prep}.

Taking into account the likely corrections to scaling,
our data for $(q,v) = (4,v_c)$, $(5,v'_c)$ and $(q_c,-1)$
are compatible with all three models
being in the 4-state Potts universality class.

\paragraph{Conclusion.}
$\!\!\!$We have given:
(a) an analytical existence
argument for a finite-temperature phase transition
in a class of 4-state Potts antiferromagnets;
(b) a prediction of the universality class; 
(c) large-scale numerics, using two complementary techniques,
to determine critical exponents;
(d) determination of $q_0$ and $q_c$ as well as $v_c$;
and (e) the surprising prediction of a finite-temperature phase transition
also for $q=5$ on the BH lattice \cite{all_of_us_in_prep}.

\begin{acknowledgments}
This work was supported in part by
NSFC grant 10975127, the Chinese Academy of Sciences,
French grant ANR-10-BLAN-0414, the Institut Universitaire de France,
Spanish MEC grants FPA2009-08785 and MTM2008-03020,
and NSF grant PHY--0424082.
\end{acknowledgments}

\vspace*{-2mm}



\begin{thebibliography}{99}
\bibitem{note_TRI_q=2}  On the triangular lattice, both $q=2$ and $q=4$ 
are critical at zero temperature and have height representations
\cite{Blote_82_etc} \cite{Moore_00}, but $q_c = 4$.
\bibitem{note_criticality}  This suggestion could possibly be made
rigorous if \reff{eq.prop_4color} could be generalized to an identity
for suitable correlation functions.
\bibitem{note_Xt}  The 4-state Potts value $X_t = \smfrac{1}{2}$
is more relevant than the Ising value $X_t = 1$ and hence dominates.
\bibitem{note_m=4}  The $m$th minimal model \cite{DiFrancesco_97}
has central charge $c = 1 - 6/[m(m+1)]$
and critical exponents $X_{r,s} = [(r(m+1)-sm)^2 - 1]/[2m(m+1)]$
for integer (and sometimes half-integer \cite{Jacobsen_09}) $r,s$.
It is the continuum limit both of the $q = B_{m+1}$ critical Potts ferromagnet
and of the $q = B_m$ tricritical Potts ferromagnet.
\end{thebibliography}
\end{document}